\documentclass[journal]{IEEEtran}

\usepackage[cmex10]{amsmath}
\usepackage{algorithmic}
\usepackage{array}

\usepackage{epsf}
\usepackage{epsfig}
\usepackage{amsmath}
\usepackage{amsfonts}
\usepackage{amssymb}
\usepackage{amsxtra}
\usepackage{latexsym}
\usepackage{soul,color}
\usepackage{graphics}
\usepackage{graphicx}
\usepackage{verbatim}   
\usepackage{epstopdf}
\usepackage[caption=false,font=footnotesize]{subfig}

\usepackage{algorithm}
\usepackage{algorithmic}
\usepackage{mathrsfs}
\usepackage{cite}
\usepackage[cmex10]{amsmath}
\usepackage{algorithmic}
\usepackage{array}
\usepackage{mdwmath}
\usepackage{mdwtab}
\usepackage{eqparbox}
\usepackage{url}
\usepackage{float}
\usepackage{etoolbox} 
\title{Model Based Iterative Reconstruction With Spatially Adaptive Sinogram Weights for Wide-Cone Cardiac CT}
\author{ 
Amirkoushyar~Ziabari~\IEEEmembership{, member,~IEEE},
Dong Hye Ye~\IEEEmembership{, member,~IEEE}, 
Lin Fu,
Somesh Srivastava~\IEEEmembership{, member,~IEEE}, 
Ken D. Sauer~\IEEEmembership{, member,~IEEE}, 
Jean-Baptiste Thibault~\IEEEmembership{, member,~IEEE}, and 
Charles A. Bouman~\IEEEmembership{, fellow,~IEEE} 

\thanks{A.~Ziabari, D.~Ye, and C.~A.~Bouman (corresponding author) are with the ECE Department, Purdue University, IN 47907 USA (Emails: aziabari@purdue.edu; yed@purdue.edu; Bouman@purdue.edu.)}
\thanks{S.~Srivastava and J.-B.~Thibault are with GE Healthcare Technologies, Waukesha, WI 53188 USA (Emails: someshsrivastava@ge.com; Jean-Baptiste.Thibault@med.ge.com)}
\thanks{L.~Fu is with GE Global Research, Niskayuna, NY, 12309 USA (Email: fulin@ge.com)}
\thanks{K.~D.~Sauer is with the EE Department, University of Notre Dame, Notre Dame, IN 46556-5637 USA (Email:sauer@nd.edu).}
\vspace{-1cm}
}

\begin{document}
\maketitle
\begin{abstract}
With the recent introduction of CT scanners with large cone angles, wide coverage detectors now provide a desirable scanning platform for cardiac CT that allows whole heart imaging in a single rotation. On these scanners, while half-scan data is strictly sufficient to produce images with the best temporal resolution, acquiring a full 360 degree rotation worth of data is beneficial for wide-cone image reconstruction at negligible additional radiation dose. 
Applying Model-Based Iterative Reconstruction (MBIR) algorithm to the heart has already been shown to yield significant enhancement in image quality for cardiac CT. 
But imaging the heart in large cone angle geometry leads to apparently conflicting data usage considerations. On the one hand, in addition to using the fastest available scanner rotation speed, a minimal complete data set of 180 degrees plus the fan angle is typically used to minimize both cardiac and respiratory motion. 
On the other hand, a full 360 degree acquisition helps better handle the challenges of missing frequencies and incomplete projections associated with wide-cone half-scan data acquisition. 
In this paper, we develop a Spatially Adaptive sinogram Weights MBIR algorithm, deemed SAW-MBIR, that is designed to achieve the benefits of both half-scan and full-scan reconstructions in order to maximize temporal resolution performance over the heart region while providing stable results over the whole volume covered with the wide-area detector. 
Spatially-adaptive sinogram weights applied to each projection measurement in SAW-MBIR are designed to selectively perform back projection from the full-scan and half-scan portion of the sinogram based on both projection angle and reconstructed voxel location.
We demonstrate with experimental results of SAW-MBIR applied to whole-heart cardiac CT clinical data that overall temporal resolution performance matches half-scan results while full volume image quality compares positively to the standard MBIR reconstruction of full-scan data.
\end{abstract}

\vspace{-.4cm}
\section{Introduction}
{
\IEEEPARstart{C}{ardiac} CT reconstruction requires high temporal resolution to capture the moving heart. 
To achieve high temporal resolution in the reconstructed image, it is desirable to reconstruct from limited-view angle projections, i.e. a minimal complete dataset (180 + the fan angle), also referred to as half-scan, instead of a full-scan acquisition with 360 degrees of data~\cite{Nuyts2013,Hsieh2013}. 
Although half-scan is sufficient for image reconstruction in the mid-plane, it results in incomplete projections and missing spatial frequencies for image planes under larger cone angles.
Without explicit corrections, analytical algorithms such as filtered back projection (FBP) may lead to artifacts when performing reconstruction from half-scan data at high cone angles~\cite{Pack2013,Tang2010}. 

On the other hand, wide detector apertures and large cone angles are advantageous in cardiac CT as they allow the acquisition of the whole heart in a single rotation, with reduced overall scan time over the target volume and higher X-ray tube efficiency~\cite{maass2010comparing}. 
While most conventional clinical CT platforms with smaller detectors rely only on half-scan data for heart imaging, wide-area detectors allow full-scan acquisitions with 360 degree projections with negligible additional radiation dose.
In this geometry, spatially-dependent sinogram weighting can help conserve temporal resolution performance over the heart region. In analytical methods such as FBP, this is relatively straightforward because the reconstruction at each spatial location can be carried out in a closed form equation, independent of other spatial locations, as long as the weights are calculated to sum to a constant over redundant projection data contributing to the same location. 
However, large cone angles result in missing data and inconsistent projections which can lead to image distortions without explicit compensation~\cite{maass2010comparing}.

Model Based Iterative Reconstruction (MBIR) affords additional flexibility in data handling and has been shown to perform better than analytical methods in terms of noise, artifact reduction, and spatial resolution performance~\cite{Thibault2007,Yu2011,Ye2017}, thus allowing significant reduction in patient radiation dose~\cite{Nuyts2013,Ye2017,fuchs2014coronary}. 
But specifically for cardiac imaging, MBIR needs to balance temporal resolution performance with image spatial resolution, noise, and uniformity. 
Spatially-dependent sinogram weighting can be employed to control the contributions from full-scan and half-scan data sets depending on voxel location in MBIR as well, as long as the algorithm provides for consistent problem formulation that allows stable convergence to the minimum of the cost function.

Individual voxel locations or sub-regions of the cardiac volume may be properly reconstructed with spatially-dependent sinogram weighting in MBIR when back projecting a single full-scan wide-cone dataset which treats differently locations within the heart area vs. outside the heart. 
Parker weights~\cite{wesarg2002parker} can for instance be used for back projection of half-scan data over the heart region, whereas regions outside the heart are back projected using the corresponding full-scan sinogram weights.
In this case, the algorithm needs to deal with an unmatched forward/backward projection pair (or dual system matrix)~\cite{Kamphuis1998}. 
Such approaches have previously been investigated to reduce artifacts~\cite{Zeng1992} as well as to accelerate the convergence of IR algorithms~\cite{Kamphuis1998, Wang2015a, Zhang2016}. 
However, in these past works, the back projection operator does not contain any specific information about the locations of both the measurements and the reconstructed voxels.

In this paper, we propose the SAW-MBIR algorithm that uses spatially-adaptive sinogram weights to perform selective back projection from different sub-regions or voxels in the reconstructed image, and apply it to wide-cone angle cardiac CT.
In a single iteration, SAW-MBIR selectively performs half-scan back projection over the heart region and full-scan back projection over the rest of the volume in order to address the challenges normally associated with incomplete and missing data from the half scan geometry in regions with high cone angles. 
Experimental results demonstrate that the SAW-MBIR algorithm achieves consistent temporal resolution with half scan MBIR reconstruction over the heart region, and good image quality consistent with full-scan MBIR reconstruction in the rest of the reconstructed volume, all in a consistent algorithm framework operating with a single reconstruction. 

\section{SAW-MBIR}\label{sec2}

\subsection{Theoretical Formulation}

The objective of this work is to develop an MBIR formulation with spatially-adaptive sinogram weighting (SAW-MBIR) for selective back projection of cardiac full-scan CT data. 
In this approach, separate back projections are performed from the sinogram residual corresponding to different sub-regions in the reconstructed image volume. 
Subsequently, the back projection results are weighted using a mask applied to the spatial location of each sub-regions.
When used over half-scan and full-scan projection regions, this supports the reconstruction of an image volume with good temporal resolution over the central region while using full sampling to reconstruct the rest of the volume.

To explain the idea, we use gradient descent to find the solution to the problem:
\begin{equation}
y = Ax 
\label{eq1}
\end{equation}
where $A \in \Re^{M\times N}$ is the system matrix, $x \in \Re^{N}$ is the unknown vector of the image, and $y \in \Re ^{M}$ is the vector of sinogram measurements. 
The corresponding maximum {\it a posteriori} (MAP) cost function is:
\begin{equation}
f(x) = \frac{1}{2} ||y-Ax||_W^2 + \Phi(x)
\label{eq2}
\end{equation}

The norm in the first term is the data fidelity term, where $W$ is a diagonal statistical weighting matrix that models the noise~\cite{Ye2017} and $\Phi(x)$ is the regularization function.
We can then write the gradient descent update equation as follows:
\begin{equation}
g(x) = \nabla f(x) = A^T W (Ax-y) + \nabla\Phi(x)
\label{eq3}
\end{equation}
\begin{equation}
x^{(k+1)} = x^k - \alpha g(x),
\label{eq4}
\end{equation}
where $k$ denotes the iteration number and $\alpha$ is the step size. 
Here, $A^T$ is the matched back projection operator to the forward projection system matrix $A$.

For a cardiac CT scan that includes projections from the full 360 degree rotation scan, we denote $y$ the complete set of projections (full-scan), and consider the limited view angle measurements $y_{h}$ (half-scan) as a subset of the full-scan measurements.
We note $y_{h'}$ the complement of $y_{h}$, which includes all the rest of the projections from outside the limited-view angles. 
We define:
\begin{equation}
    y \equiv 
    \begin{bmatrix}
           y_{h} \\
           y_{h'} 
    \end{bmatrix}
\label{eq5}
\end{equation}

We also partition the image $x$ into two regions of $x_{m}$, which corresponds to the target region (here, the heart region) and should be back projected from half-scan measurement data, and $x_{m'}$, which corresponds to the background region.
\begin{equation}
    x \equiv 
    \begin{bmatrix}
           x_{m} \\
           x_{m'} 
    \end{bmatrix}
\label{eq6}
\end{equation}

Consistently, we partition the $A$ matrix into four sub-matrices:
\begin{equation}
    A \equiv 
    \begin{bmatrix}
           A_{hm}  & A_{hm'}\\
           A_{h'm} & A_{h'm'}
    \end{bmatrix}
\label{eq7}
\end{equation}
$A_{hm}$ and $A_{hm'}$ are the sub-matrices that project $x_m$ and $x_m'$ into the half-scan measured sinogram $y_h$, respectively. On the other hand, $A_{h'm}$ and $A_{h'm'}$ project $x_m$ and $x_m'$ into the region outside the half-scan measured sinogram $y_h'$.

We can rewrite equation (\ref{eq1}) as follows: 
\begin{equation}
    \begin{bmatrix}
           y_{h} \\
           y_{h'} 
    \end{bmatrix}
    = 
    \begin{bmatrix}
           A_{hm}  & A_{hm'}\\
           A_{h'm} & A_{h'm'}
    \end{bmatrix}
    \begin{bmatrix}
           x_{m} \\
           x_{m'} 
    \end{bmatrix}
\label{eq9}
\end{equation}
Using half-scan projections is advantageous in improving the temporal resolution relative to the full-scan data.
The corresponding back projection matrix can be written as:
\begin{equation}
    A_{half}^T \equiv 
    \begin{bmatrix}
           A_{hm}^T  & 0 \\
           A_{hm'}^T & 0
    \end{bmatrix}
\label{eq10a}
\end{equation}

It is challenging to reconstruct image slices at high cone angles from only half-scan data because of incomplete and missing projections in the half-scan dataset. Instead, one may benefit from using full-scan projections outside the central primary region of interest, in conjunction with half-scan data only for the center region, in order to both improve temporal resolution at the center and maintain high image quality throughout.

\vspace{-0.2cm}
 \begin{algorithm}
\caption{SAW-MBIR \label{alg1}}
\begin{algorithmic}
 \scriptsize
 \STATE $ y \leftarrow $ measured sinogram
 \STATE $ x^0 \leftarrow $ FBP 
 \STATE $ \alpha \leftarrow $ step size 
 \STATE For  $k$  iterations  \{ 
 \STATE $ g_{s}(x^k) = A_{masked}^TW(Ax^k-y) + \nabla\Phi(x^k) $
 \STATE $ x^{(k+1)} = x^k - \alpha g_{s}(x^k)$ \}
\end{algorithmic}
\end{algorithm}
\vspace{-0.2cm}

Here, we introduce the SAW-MBIR algorithm that uses spatially-adaptive sinogram weights to perform selective back projection while retaining a consistent framework for iterative optimization. We first define the masked back projection operator $A_{masked}^T$:
\begin{equation}
    A_{masked}^T \equiv 
    \begin{bmatrix}
           A_{hm}^T  & 0 \\
           A_{hm'}^T & A_{h'm'}^T
    \end{bmatrix}
\label{eq10b}
\end{equation}
Compared to $A^T$, by setting $A_{h'm} = 0$, $A_{masked}^T$ decouples $y_{h'}$ from $x_m$ when back projecting the sinogram residual to the image.
The back projector is masked (i.e. set to zero) depending on the spatial location of the regions that are outside the half-scan projection data.
Putting $A_{masked}^T$ in equation (\ref{eq3}), we get the following pseudo-gradient:
\begin{equation}
\begin{split}
g(x) = g_{s}(x) \equiv 
    \begin{bmatrix}
           [g_{s}(x)]_{m} \\
           [g_{s}(x)]_{m'}
    \end{bmatrix}
			+ \nabla\Phi(x) \\
            = A_{masked}^TW(Ax-y) + \nabla\Phi(x) ,
\end{split}
\label{eq11}
\end{equation}
where:
\begin{equation}
    [g_{s}(x)]_{m} \equiv 
    	\begin{bmatrix}
           A_{hm} \\
           0
         \end{bmatrix}
         ^T
         W(Ax-y)
\label{eq12}
\end{equation}
\begin{equation}
    [g_{s}(x)]_{m'} \equiv 
    	\begin{bmatrix}
           A_{hm'} \\
           A_{h'm'} 
         \end{bmatrix}
         ^T
         W(Ax-y)
\label{eq13}
\end{equation}

And the update equation becomes:
\begin{equation}
x^{(k+1)} = x^k - \alpha g_{s}(x^k)
\label{eq14}
\end{equation}

Eqs. (\ref{eq12}) and (\ref{eq13}) above show how the proposed SAW-MBIR algorithm selectively performs half-scan back projection of locations within the mask (Eq. (\ref{eq12})) and full-scan back projection of locations outside the mask (Eq. (\ref{eq13})). 
A summary of the proposed method is shown in algorithm \ref{alg1}.

\vspace{-0.5cm}
\subsection{Application to Wide-Cone Cardiac CT}
SAW-MBIR can be applied in whole-heart wide-cone cardiac CT to produce better results than standard MBIR.
An illustration of the transverse view of the cardiac CT data acquisition is shown in Fig.~\ref{fig1}a.
Cone-beam projections are drawn for two opposite view angles. 
Standard MBIR with full-scan projection data can produce good image quality in the full volume by reducing noise and artifacts compared to analytical methods, but does not inherently produce the temporal resolution of the half-scan acquisition over the heart area typically covered in the fully-sampled (i.e. light purple) region.
The SAW-MBIR algorithm introduced in section II.A. can be used to improve the temporal resolution in the heart region relative to full-scan. 

\begin{figure}[ht]
\centering
\vspace{-.2cm}
\includegraphics [scale=.23]{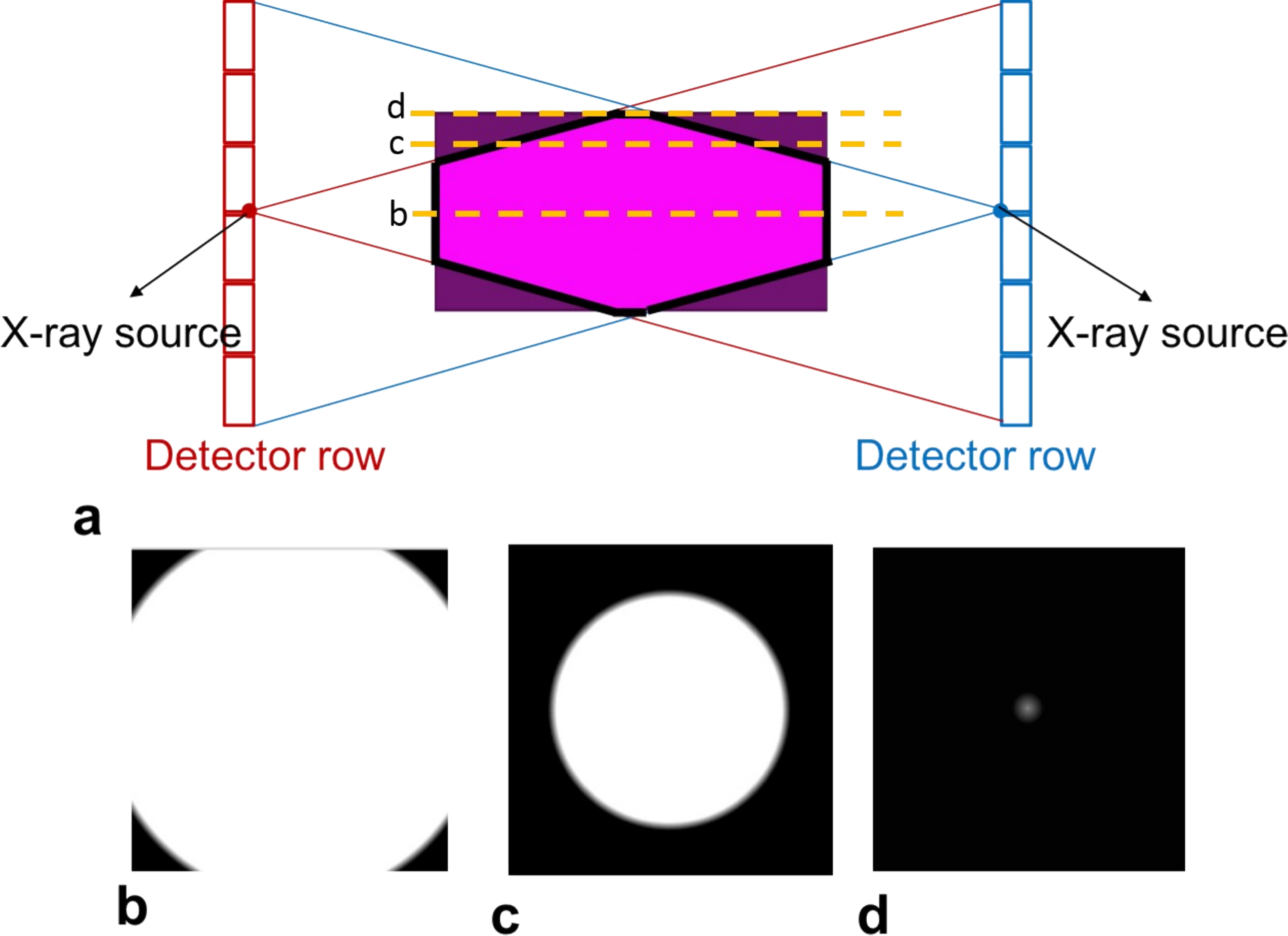}
\caption{
a. A schematic of the transverse view of the cardiac CT data acquisition. 
Projections from two opposite view angles are drawn in red and blue.
The desired reconstruction volume is indicated by the rectangular box at the center.
A mask is defined to differentiate between half-scan and full-scan back projected regions.
Our intention is to reconstruct the light purple region primarily based on the half-scan data,
and the dark purple region primarily based on the full-scan data.
Slices of the reconstruction mask are shown: 
(b) at the center, 
(c) between center and edge, and 
(d) at the edge.
}
\label{fig1}
\vspace{-.4cm}
\end{figure}

We define a mask to distinguish between the area where complete projection data is available from the half-scan data and the rest of the volume. 
We intend to reconstruct the light purple region of Fig.~\ref{fig1}a primarily from the half-scan data, and the dark purple region primarily from the full-scan data.
Cross sections of the mask taken at the center, between the center and the edge, and at the edge, are shown in Fig.~\ref{fig1}b, c, and d, respectively.

The mask is set to one based on the spatial location of the voxels with respect to the light and dark purple regions of Fig.~\ref{fig1}a.
Since the half-scan and full-scan back projection operators only differ in the angular range of integration, the mask is used to implement $A_{masked}^T$ of Eq. (\ref{eq10b}).
The masked back projector selectively performs back projection from the half-scan and full-scan measurements at each iteration of MBIR.
Basically, the masked back projector works as if only measurements inside the region corresponding to the half-scan data are back projected by half-scan back projector of Eq. (\ref{eq12}), and the rest of the measurements are back projected using the full-scan back projector of Eq. (\ref{eq13}).

\section{Results and Discussions}
\label{sec3}

Here we use the SAW-MBIR algorithm to improve upon the standard MBIR implementation in ~\cite{Fu2013,Yu2013} applied to cardiac CT.
This approach uses a preconditioned gradient-based IR algorithm to simultaneously update all the voxels. 
Further, the ordered subset (OS) method~\cite{hudson1994accelerated} is used to calculate the sub-gradient for each subset at each iteration, then a preconditioner operator is used to accelerate the high frequency convergence, and the Nesterov's method~\cite{nesterov2012make} further reduces the number of iterations to achieve convergence.
The method included a line-search step that ensures the monotonic decrease of the cost function relative to the previous estimate.

It is worth noting that normally making the statistical weighting dependent on location in the image volume would result in an inconsistent problem formulation for iterative convergence.
Using a line search, however, mitigates the concern of inconsistent weights, and  in turn guarantees convergence and stability. The convergence path would be influenced by both the gradient of the original cost function~\ref{eq3} and the pseudo-gradient~\ref{eq11}.
Intuitively, a fixed point of the iteration is reached when the two gradient vectors become orthogonal or have a negative inner product. We leave further theoretical analysis of the properties of the convergence point to future studies, so that this paper would focus on achieving the intended image quality benefits for wide-cone cardiac CT.

Clinical datasets from GE Healthcare Revolution CT scanner with 160~mm detector aperture at the isocenter are used here, with a single axial rotation covering the whole heart. 
Comparisons between full-scan and half-scan MBIR
as well as the proposed SAW-MBIR algorithm are shown in Figs.~\ref{fig2} and ~\ref{fig3}.

Fig.~\ref{fig2} compares the performance of SAW-MBIR against full-scan and half-scan MBIR for wide-cone cardiac CT reconstruction in the chest (Figs.~\ref{fig2}a-c), and in the liver (Figs.~\ref{fig2}d-f) regions. 
Figs.~\ref{fig2}a-c correspond to a slice between cross sections c and d (closer to d at the edge) in Fig.~\ref{fig1}, and Figs.~\ref{fig2}d-f belong to a same region on the opposite side of the reconstruction volume.

The full-scan and half-scan MBIR results are shown in Figs.~\ref{fig2}a, and d, and Figs.~\ref{fig2}b, and e, respectively. 
Figs.~\ref{fig2}c, and f belong to SAW-MBIR results.
SAW-MBIR perform consistently as good as full-scan MBIR outside the heart region. 
Half-scan MBIR, on the other hand, shows some distortions in the regions outside the heart due to incomplete and missing projection data.

\begin{figure}[ht]
\centering
\includegraphics [scale=.24]{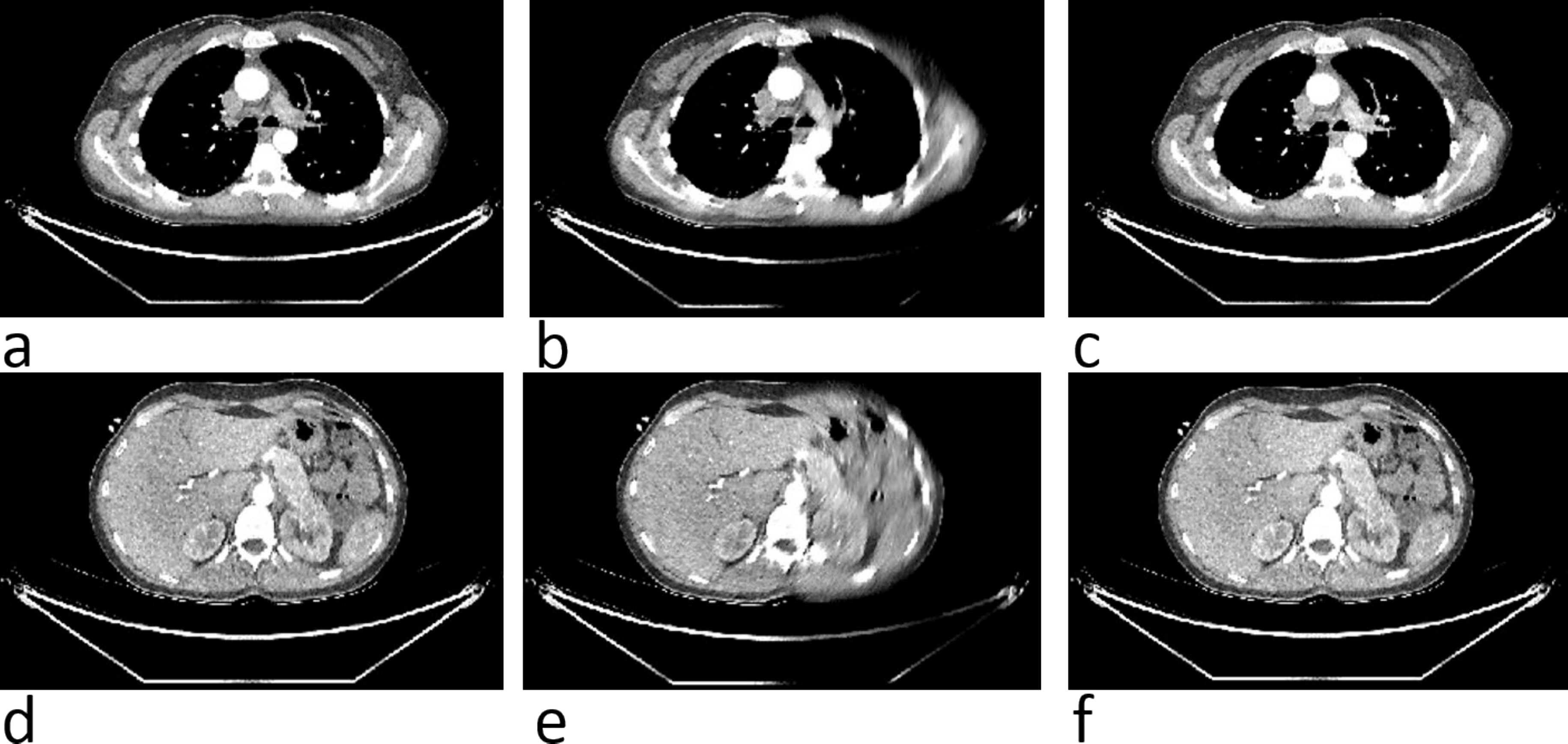}
\caption{
Comparison between reconstructed cardiac CT images in the chest and liver regions.
a, and d. Full-scan MBIR.
b, and e. Half-scan MBIR. 
c, and f. SAW-MBIR. 
Window settings show the [-200 200]~HU range.
SAW-MBIR shows consistent image quality with full-scan outside the heart where half-scan MBIR is distorted.
Panels a-c correspond to a slice between cross sections c and d (closer to d at the edge) in Fig. 1, and panels d-f are from a slice on the opposite side in the reconstructed volume. 
}
\label{fig2}
\end{figure}

\begin{figure}[ht]
\centering
\includegraphics [scale=.35]{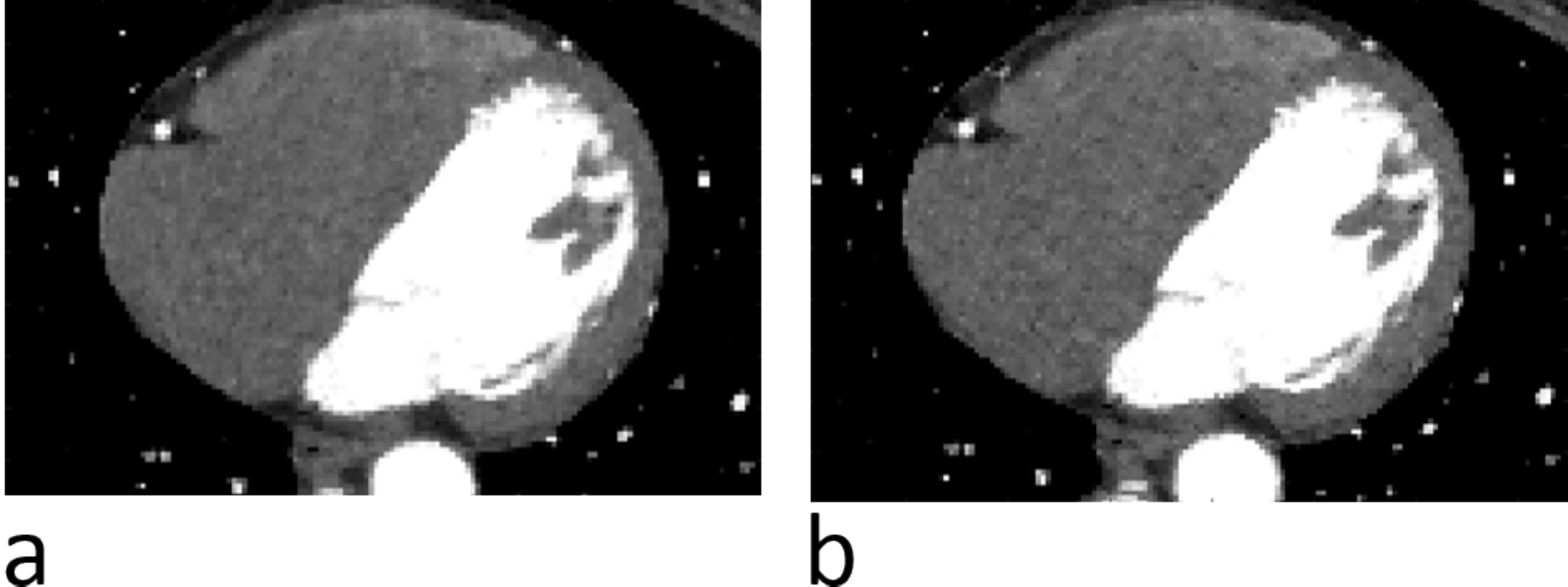}
\caption{Impact on temporal resolution.
An expanded view of the heart region in the reconstructed center slice of the wide-cone acquisition is shown in 
(a)  half-scan MBIR, and 
(b) SAW-MBIR. 
Window settings show the [-200 500]~HU range.}
\label{fig3}
\end{figure}

\begin{figure}[ht]
\centering
\includegraphics [scale=.3]{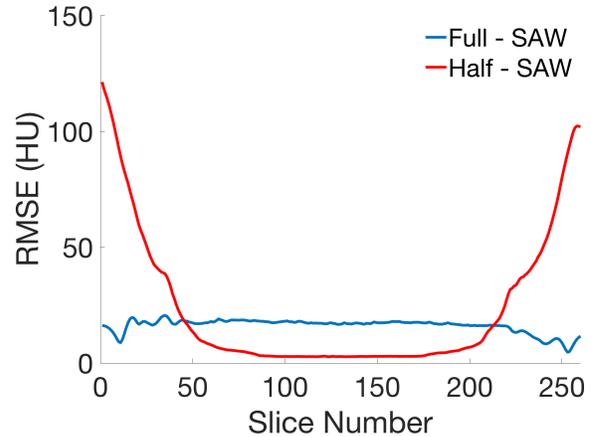}
\caption{RMSE at each slice.
The RMSE of full-scan MBIR (blue) and half-scan MBIR (red) with respect to SAW MBIR is illustrated. 
Background pixels with zero intensity were excluded.
}
\label{fig4}
\end{figure}

We also compared reconstruction results specifically in the heart region (center slices) between half-scan, and SAW-MBIR in Figs.~\ref{fig3}a, and b. 
Qualitative comparisons show close agreement between SAW-MBIR and half-scan results.

Further, the root mean squared error (RMSE) between full-scan MBIR and SAW MBIR, as well as between half-scan MBIR and SAW-MBIR at each slice location is computed.
The results are shown in Fig.~\ref{fig4}.
It is clear that in the non-heart regions (the edge slices) SAW-MBIR matches better with full-scan MBIR without the distortions of half-scan MBIR. 
The results also suggest very good agreement between the half-scan and SAW-MBIR results in the heart region, which further verify that the SAW-MBIR maintains comparable temporal resolution as the half-scan MBIR.

It is worth noting that, to reduce half-scan artifacts while maintaining temporal resolution, Cho in ~\cite{cho2014improving} proposed akin approach where they use extra measurements (such as full-scan data) to modify the statistical weighting using an extrapolation scheme. 
However, the experimental results on numerical phantoms proved challenging to tune parameters of their model to obtain optimum solution with temporal resolution of half-scan and reduced artifacts as good as full-scan results.

\section{Conclusion}
\label{sec4}
In this work, we developed the SAW-MBIR algorithm that uses spatially-adaptive weights to perform selective back projection of sinogram residuals to different sub-regions or voxels in the reconstructed image.
Back projection weights may be determined based on both the measurement position and the location of the reconstructed voxel in the field of view.
We examined the performance of the SAW-MBIR using whole-heart cardiac CT clinical data sets with temporal heart motion. 
The experimental results obtained using the SAW-MBIR demonstrate marked performance in achieving high temporal resolution in the heart region with similar image quality to standard full-scan MBIR outside the heart region.
While cardiac CT is shown as an example, the method can be extended to other scan geometries or imaging modalities wherein image artifacts or image degradations may be spatially localized, for instance with scatter, low signal, or metal artifacts.  
\bibliographystyle{IEEEbib}
\bibliography{Refs_ISBI}

\end{document}